%% file: main.tex
\documentclass[a4paper,11pt]{article}
\usepackage{pos}
\usepackage{amsmath}
\usepackage{mathtools}
\usepackage{mdframed}
\usepackage{microtype}

\renewcommand{\logo}{\relax}

\title{Theory updates for parton distributions in Hessian formalism}

\author*[a]{Petja Paakkinen}
\author[b,c]{Hannu Paukkunen}
\author[b,c]{Sami Yrjänheikki}

\affiliation[a]{CERN, Theoretical Physics Department,\\
1211 Geneva 23, Switzerland}
\affiliation[b]{University of Jyväskylä, Department of Physics,\\
P.O. Box 35, 40014 University of Jyväskylä, Finland}
\affiliation[c]{Helsinki Institute of Physics,\\
P.O. Box 64, 00014 University of Helsinki, Finland}

\emailAdd{petja.paakkinen@cern.ch}
\emailAdd{hannu.t.paukkunen@jyu.fi}
\emailAdd{sami.a.yrjanheikki@jyu.fi}

\abstract{We present a new extension to the toolbox of parton-distribution reweighting methods, which enables a general user to study the impact of an updated theory prediction on the results of a pre-existing parton-distribution global analysis. This new method is a combination of the well-known reweighting method with a new deweighting variant where a dataset is removed from the original analysis in an approximate way. Specific use cases of this method could include e.g.\ testing the impact of updating the treatment of an observable from N$^n$LO to N$^{n+1}$LO precision in the global analysis, or including previously ignored electroweak, mass, resummation, or higher-twist effects, or even testing the impact of some beyond-standard-model physics. We discuss the implementation of this method in the Hessian formalism, and show its use in a single case study, where we update the treatment of dimuon production in neutrino-induced DIS with the novel semi-inclusive DIS approach.}

\FullConference{The 33rd International Workshop on Deep Inelastic Scattering and Related Subjects (DIS2026)\\
4 - 8 May 2026\\
Bologna, Italy\\}

\begin{document}

\include{defs}

\maketitle

\section{Introduction}

Reweighting methods in different frameworks (Monte Carlo and Hessian) have nowadays been established as standard tools for obtaining estimates of updated parton distribution functions (PDFs) after including new data on top of an existing global analysis. We present here an extension to the method in the Hessian PDF formalism~\cite{Paukkunen:2014zia} that also allows removing a dataset from the global analysis or updating the theory description of a certain process. As an application, we use this method to update the treatment of neutrino-induced dimuon production in deep inelastic scattering (DIS) in PDF extraction with the novel semi-inclusive DIS (SIDIS) approach~\cite{Helenius:2024fow}.

\section{Recap: PDF reweighting in Hessian formalism}

The Hessian formalism for PDF uncertainties~\cite{Pumplin:2001ct} relies on expanding the $\chi^2$ function in terms of the PDF parameters $\mathbf{a}$. Keeping only the leading quadratic behaviour, one can write the global-analysis figure of merit $\chi^2_\mathtt{GA}$ around best-fit parameters $\mathbf{a}_\mathtt{min,GA}$ as
\begin{equation}
\begin{split}
    \chi^2_\mathtt{GA}(\mathbf{a}) &= (\mathbf{D} - \mathbf{T}(\mathbf{a}))^T \, {C}^{-1} \, ({\mathbf{D}} - {\mathbf{T}}(\mathbf{a})) \\
    &\approx \chi^2_\mathtt{min,GA} + (\mathbf{a} - \mathbf{a}_\mathtt{min,GA})^T \, \Hga \, (\mathbf{a} - \mathbf{a}_\mathtt{min,GA}) \\
    &= \chi^2_\mathtt{min,GA} + \mathbf{z}^2 \, ,
\end{split}
\label{eq:chi2ga}
\end{equation}
where $\mathbf{D}$ are the fitted data with covariance matrix $C$ and $\mathbf{T}(\mathbf{a})$ the corresponding theory predictions, and on the last line we transformed into whitened parameters, $\mathbf{z} = \Wga (\mathbf{a} - \mathbf{a}_\mathtt{min,GA})$ such that the Hessian matrix diagonalizes with $\Hga = \WgaTransp \Wga$, obtained e.g.\ through eigenvalue decomposition. The Hessian PDF central and error sets $S_{\mathtt{GA},0}, S_{\mathtt{GA},k}^\pm$ are given (to this order) in terms of
\begin{equation}
\begin{split}
  f_i[S_{\mathtt{GA},0}] &= f_i(\mathbf{z}[S_{\mathtt{GA},0}]), \qquad z_j[S_{\mathtt{GA},0}] = 0\ \text{for all}\ j \, , \\
  f_i[S_{\mathtt{GA},k}^\pm] &= f_i(\mathbf{z}[S_{\mathtt{GA},k}^\pm]), \qquad z_j[S_{\mathtt{GA},k}^\pm] = \pm\,\delta_{jk}\sqrt{\Delta\chi^2} \, ,
\end{split}
\end{equation}
where $i$ labels the parton flavour, $j$, $k$ the parameter eigendirections and $\Delta\chi^2$ is the tolerance given for the allowed $\chi^2$ growth.
The PDF uncertainty can be expressed e.g.\ through the symmetric error prescription
\begin{equation}
  \Delta f_i = \frac{1}{2}\sqrt{\sum_k (f_i[S_k^+] - f_i[S_k^-])^2}.
  \label{eq:symerr}
\end{equation}

In the PDF reweighting, we want to study the impact of adding a new dataset, with the figure of merit becoming
\begin{equation}
\begin{split}
  \chi^2_\mathtt{rw}(\mathbf{z}) &= \chi^2_\mathtt{GA}(\mathbf{z}) + (\Dnew - \Tnew(\mathbf{z}))^T \, \CnewInv \, (\Dnew - \Tnew(\mathbf{z})) \\
  &\approx \chi^2_\mathtt{min,GA} + \mathbf{z}^2 + (\Dnew - \TnewZero - \Jnew\mathbf{z})^T \, \CnewInv \, (\Dnew - \TnewZero - \Jnew\mathbf{z}) \\
  &= \chi^2_\mathtt{min,rw} + \tilde{\mathbf{z}}^2
\end{split}
\end{equation}
where we used a linear expansion $\Tnew(\mathbf{z}) \approx \TnewZero + \Jnew\mathbf{z}$ in addition to Eq.~\eqref{eq:chi2ga} and $\tilde{\mathbf{z}} = \Wrw (\mathbf{z} - \mathbf{z}_\mathtt{min,rw})$ with $\Wrw$ obtained again through a suitable Hessian matrix diagonalization. This yields a simple analytical solution for the reweighted PDFs~\cite{Paukkunen:2014zia}:
\begin{mdframed}
  The new minimum after reweighting is at
  \begin{equation}
    \mathbf{z}_\mathtt{min,rw} = \HrwInv \, \JnewTransp \, \CnewInv \, (\Dnew - \TnewZero) \phantom{\big]}
    \label{eq:zminrw}
  \end{equation}
  and the $\mathbf{z}$-space Hessian is
  \begin{equation}
    \Hrw = 1 + \JnewTransp \, \CnewInv \, \Jnew \, .
    \label{eq:Hrw}
  \end{equation}
\end{mdframed}
The reweighted central and error sets are given as
\begin{equation}
\begin{split}
  f_i[S_{\mathtt{rw},0}] &= f_i[S_{\mathtt{GA},0}] + {\textstyle\sum_j} \, w_j[S_{\mathtt{rw},0}] \, {\textstyle\frac{1}{2}}(f_i[S_{\mathtt{GA},j}^+] - f_i[S_{\mathtt{GA},j}^-]) \, , \\
  f_i[S_{\mathtt{rw},k}^\pm] &= f_i[S_{\mathtt{GA},0}] + {\textstyle\sum_j} \, w_j[S_{\mathtt{rw},k}^\pm] \, {\textstyle\frac{1}{2}}(f_i[S_{\mathtt{GA},j}^+] - f_i[S_{\mathtt{GA},j}^-]) \, ,
  \label{eq:rwpdfs}
\end{split}
\end{equation}
where the weight factors are
\begin{equation}
\begin{split}
  w_j[S_{\mathtt{rw},0}] &= {\textstyle\frac{1}{\sqrt{\Delta\chi^2}}} (\mathbf{z}_\mathtt{min,rw})_j \, , \\
  w_j[S_{\mathtt{rw},k}^\pm] &= {\textstyle\frac{1}{\sqrt{\Delta\chi^2}}} (\mathbf{z}_\mathtt{min,rw})_j \pm (\WrwInv)_{jk} \, .
  \label{eq:weights}
\end{split}
\end{equation}

\section{Novelty: PDF deweighting and updating in Hessian formalism}

Let us study the case where we want to find the impact of \emph{removing} an old dataset from the global analysis. We call this method deweighting, with the new $\chi^2$ function being
\begin{equation}
\begin{split}
  \chi^2_\mathtt{dw}(\mathbf{z}) &= \chi^2_\mathtt{GA}(\mathbf{z}) - (\Dold - \Told(\mathbf{z}))^T \, \ColdInv \, (\Dold - \Told(\mathbf{z})) \\
  &\approx \chi^2_\mathtt{min,GA} + \mathbf{z}^2 - (\Dold - \ToldZero - \Jold\mathbf{z})^T \, \ColdInv \, (\Dold - \ToldZero - \Jold\mathbf{z}) \\
  &= \chi^2_\mathtt{min,dw} + \tilde{\mathbf{z}}^2
\end{split}
\end{equation}
where now $\Told(\mathbf{z})\approx \ToldZero + \Jold\mathbf{z}$ and $\tilde{\mathbf{z}} = \Wdw (\mathbf{z} - \mathbf{z}_\mathtt{min,dw})$. This yields:
\begin{mdframed}
  The new minimum after deweighting is at
  \begin{equation}
    \mathbf{z}_\mathtt{min,dw} = \HdwInv \, \big[ - \JoldTransp \, \ColdInv \, (\Dold - \ToldZero) \big]
  \end{equation}
  and the $\mathbf{z}$-space Hessian is
  \begin{equation}
    \Hdw = 1 - \JoldTransp \, \ColdInv \, \Jold
  \end{equation}
\end{mdframed}
One can readily observe that these equations differ from Eqs.~\eqref{eq:zminrw} and~\eqref{eq:Hrw} only by appropriate sign and label changes. Furthermore, the deweighted PDFs are obtained through Eqs.~\eqref{eq:rwpdfs} and~\eqref{eq:weights} with the simple substitution $\mathtt{rw} \to \mathtt{dw}$.

We can extend this method to a general case where both removing \emph{and} adding a dataset are involved. We refer to this as PDF updating, the general $\chi^2$ function being
\begin{equation}
\begin{split}
    \chi^2_\mathtt{ud}(\mathbf{z}) &= \chi^2_\mathtt{GA}(\mathbf{z}) - (\Dold - \Told(\mathbf{z}))^T \, \ColdInv \, (\Dold - \Told(\mathbf{z})) \\ &\phantom{= \chi^2_\mathtt{GA}(\mathbf{z})}\ + (\Dnew - \Tnew(\mathbf{z}))^T \, \CnewInv \, (\Dnew - \Tnew(\mathbf{z})) \\
    &\approx \chi^2_\mathtt{min,GA} + \mathbf{z}^2 - (\Dold - \ToldZero - \Jold\mathbf{z})^T \, \ColdInv \, (\Dold - \ToldZero - \Jold\mathbf{z}) \\ &\phantom{\approx \chi^2_\mathtt{min,GA} + \mathbf{z}^2}\ + (\Dnew - \TnewZero - \Jnew\mathbf{z})^T \, \CnewInv \, (\Dnew - \TnewZero - \Jnew\mathbf{z}) \\
    &= \chi^2_\mathtt{min,ud} + \tilde{\mathbf{z}}^2
\end{split}
\end{equation}
with $\tilde{\mathbf{z}} = \Wud (\mathbf{z} - \mathbf{z}_\mathtt{min,ud})$ such that $\Hud = \WudTransp \Wud$.
\begin{mdframed}
  The new minimum after updating is at
  \begin{equation}
    \mathbf{z}_\mathtt{min,ud} = \HudInv \, \big[ - \JoldTransp \, \ColdInv \, (\Dold - \ToldZero) + \JnewTransp \, \CnewInv \, (\Dnew - \TnewZero) \big]
  \end{equation}
  and the $\mathbf{z}$-space Hessian is
  \begin{equation}
    \Hud = 1 - \JoldTransp \, \ColdInv \, \Jold + \JnewTransp \, \CnewInv \, \Jnew
  \end{equation}
\end{mdframed}
The updated PDFs are again obtained from Eqs.~\eqref{eq:rwpdfs} and~\eqref{eq:weights} with a substitution $\mathtt{rw} \to \mathtt{ud}$. This general form reduces to the reweighting and deweighting expressions by taking the appropriate zero-data limits. On another limit, by setting $\Dnew \!=\! \Dold$, $\Cnew \!=\! \Cold$, one is making a \emph{theory update} on the PDFs. Possible applications involve testing the impact of updating an observable from N$^n$LO to N$^{n+1}$LO precision, including electroweak, mass, resummation, or higher-twist effects, or even testing the impact of some beyond-standard-model physics. Alternatively, taking $\Tnew \!=\! \Told$ leads to a \emph{data update}, allowing e.g.\ applying some correction or changing the correlation model.

\section{Case study: Dimuon production in neutrino-nucleus DIS}

We apply here the newly developed formalism to update the treatment of dimuon production in neutrino-nucleus DIS for the analysis of proton and nuclear PDFs. Traditionally, this process has been included in the global analyses by calculating the charm-production cross section and applying a semileptonic branching fraction and acceptance corrections to it. The idea here is to replace the old theory with the new SIDIS approach that involves data-fitted charmed-hadron fragmentation functions and a muonic decay width~\cite{Helenius:2024fow}. We take into account also the relevant heavy-quark mass effects~\cite{Helenius:2025fpy}. The results presented here are at the level of next-to-leading order in perturbative QCD, which is the precision at which these data are included in the CT18 global analysis \cite{Hou:2019efy} considered here, but can be extended to include the known corrections of one order higher~\cite{Helenius:2026uuz}.

\begin{figure}[b]
    \centering
    \includegraphics[width=\textwidth]{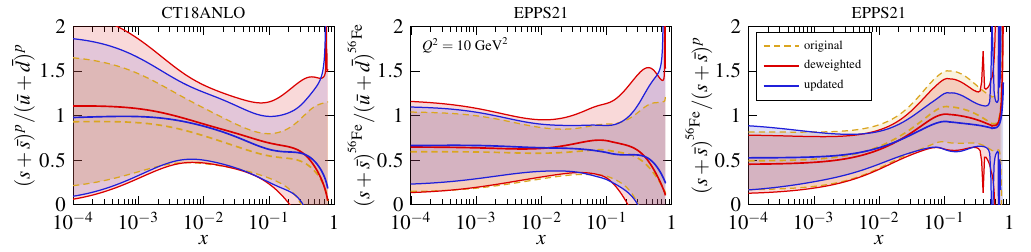}
    \caption{Preliminary results on deweighting and updating the CT18ANLO proton and EPPS21 nuclear PDFs. We show the strangeness suppression in proton (left) and iron (middle) and the strange-quark nuclear modification in iron with respect to proton (right). Original (yellow dashed) refers to the initial global-analysis PDFs, deweighted (solid red) corresponds to removing the dimuon DIS data from the CT18ANLO analysis and propagating the changes also into EPPS21 PDFs, and finally the updated (solid blue) results show the impact of putting the dimuon data back in a simultaneous reweighting of CT18ANLO and EPPS21 using the SIDIS approach. Uncertainties are calculated with the symmetric prescription of Eq.~\eqref{eq:symerr}.}
    \label{fig:dimu-dwud}
\end{figure}

The preliminary results of this exercise are given in Figure~\ref{fig:dimu-dwud}. Removing the CCFR and NuTeV dimuon data~\cite{NuTeV:2001dfo,NuTeV:2007uwm} from proton PDFs makes the strange-quark uncertainties larger, as expected. On adding the data back in a simultaneous reweighting of the CT18A proton and EPPS21 nuclear PDFs \cite{{Eskola:2021nhw}}, the nuclear strange-quark PDF becomes similar to the original one, but the proton-PDF uncertainties remain larger as the constraints now apply on its product with the nuclear modification. This gives a more faithful representation of the dimuon DIS data constraints. A more extensive discussion on the implementation and results will be given elsewhere.

\section{Summary}

We have presented an extension to the PDF reweighting method, allowing a general user to perform data-removal and theory-update operations in the Hessian formalism. As a case study, we used this method to update the treatment of dimuon DIS in a PDF fit, but we expect the extended reweighting or PDF updating method to find use in various other applications.

\acknowledgments

This work has been supported by the Magnus Ehrnrooth foundation and the Center of Excellence in Quark Matter of the Research Council of Finland, project 364194. Finnish IT Center for Science, project jyy2580, is acknowledged for computing resources.

\end{document}

%% file: defs.tex
\newcommand{\Dnew}{\mathbf{D}_{\mathtt{new}\vphantom{,0}}^{\vphantom{-}}}
\newcommand{\Tnew}{\mathbf{T}_{\mathtt{new}\vphantom{,0}}^{\vphantom{-}}}
\newcommand{\Cnew}{C_{\mathtt{new}\vphantom{,0}}^{\vphantom{-}}}
\newcommand{\CnewInv}{C_{\mathtt{new}\vphantom{,0}}^{-1}}

\newcommand{\TnewZero}{\mathbf{T}_{\mathtt{new},0}^{\vphantom{-}}}
\newcommand{\Jnew}{J_{\mathtt{new}\vphantom{,0}}^{\vphantom{-}}}
\newcommand{\JnewTransp}{J_{\mathtt{new}\vphantom{,0}}^{\vphantom{-}T}}

\newcommand{\Dold}{\mathbf{D}_{\mathtt{old}\vphantom{,0}}^{\vphantom{-}}}
\newcommand{\Told}{\mathbf{T}_{\mathtt{old}\vphantom{,0}}^{\vphantom{-}}}
\newcommand{\Cold}{C_{\mathtt{old}\vphantom{,0}}^{\vphantom{-}}}
\newcommand{\ColdInv}{C_{\mathtt{old}\vphantom{,0}}^{-1}}

\newcommand{\ToldZero}{\mathbf{T}_{\mathtt{old},0}^{\vphantom{-}}}
\newcommand{\Jold}{J_{\mathtt{old}\vphantom{,0}}^{\vphantom{-}}}
\newcommand{\JoldTransp}{J_{\mathtt{old}\vphantom{,0}}^{\vphantom{-}T}}

\newcommand{\Hga}{H_\mathtt{GA}^{\vphantom{-}}}
\newcommand{\Hrw}{H_\mathtt{rw}^{\vphantom{-}}}
\newcommand{\Hdw}{H_\mathtt{dw}^{\vphantom{-}}}
\newcommand{\Hud}{H_\mathtt{ud}^{\vphantom{-}}}

\newcommand{\HgaInv}{H_\mathtt{GA}^{-1}}
\newcommand{\HrwInv}{H_\mathtt{rw}^{-1}}
\newcommand{\HdwInv}{H_\mathtt{dw}^{-1}}
\newcommand{\HudInv}{H_\mathtt{ud}^{-1}}

\newcommand{\Lrw}{\mathbf{L}_\mathtt{rw}^{\vphantom{-}}}
\newcommand{\Ldw}{\mathbf{L}_\mathtt{dw}^{\vphantom{-}}}
\newcommand{\Lud}{\mathbf{L}_\mathtt{ud}^{\vphantom{-}}}

\newcommand{\Wga}{W_\mathtt{GA}^{\vphantom{-}}}
\newcommand{\Wrw}{W_\mathtt{rw}^{\vphantom{-}}}
\newcommand{\Wdw}{W_\mathtt{dw}^{\vphantom{-}}}
\newcommand{\Wud}{W_\mathtt{ud}^{\vphantom{-}}}

\newcommand{\WgaTransp}{W_\mathtt{GA}^{\vphantom{-}T}}
\newcommand{\WrwTransp}{W_\mathtt{rw}^{\vphantom{-}T}}
\newcommand{\WdwTransp}{W_\mathtt{dw}^{\vphantom{-}T}}
\newcommand{\WudTransp}{W_\mathtt{ud}^{\vphantom{-}T}}

\newcommand{\WgaInv}{W_\mathtt{GA}^{-1}}
\newcommand{\WrwInv}{W_\mathtt{rw}^{-1}}
\newcommand{\WdwInv}{W_\mathtt{dw}^{-1}}
\newcommand{\WudInv}{W_\mathtt{ud}^{-1}}

\newcommand{\DoldA}{\mathbf{D}_{\mathtt{old},A}^{\vphantom{-}}}
\newcommand{\ToldA}{\mathbf{T}_{\mathtt{old},A}^{\vphantom{-}}}
\newcommand{\ColdInvA}{C_{\mathtt{old},A}^{-1}}

\newcommand{\ToldAZero}{\mathbf{T}_{\mathtt{old},A,0}^{\vphantom{-}}}
\newcommand{\JoldA}{J_{\mathtt{old},A}^{\vphantom{-}}}
\newcommand{\JoldATransp}{J_{\mathtt{old},A}^{\vphantom{-}T}}

\newcommand{\JnewA}{J_{\mathtt{new},A}^{\vphantom{-}}}
\newcommand{\JnewATransp}{J_{\mathtt{new},A}^{\vphantom{-}T}}

\newcommand{\DoldB}{\mathbf{D}_{\mathtt{old},B}^{\vphantom{-}}}
\newcommand{\ToldB}{\mathbf{T}_{\mathtt{old},B}^{\vphantom{-}}}
\newcommand{\ColdInvB}{C_{\mathtt{old},B}^{-1}}

\newcommand{\ToldBZero}{\mathbf{T}_{\mathtt{old},B,0}^{\vphantom{-}}}
\newcommand{\JoldB}{J_{\mathtt{old},B}^{\vphantom{-}}}
\newcommand{\JoldBTransp}{J_{\mathtt{old},B}^{\vphantom{-}T}}

\newcommand{\JnewB}{J_{\mathtt{new},B}^{\vphantom{-}}}
\newcommand{\JnewBTransp}{J_{\mathtt{new},B}^{\vphantom{-}T}}